%% file: Turb2D_LB_vf_arXiv.tex
\documentclass[secnumarabic,amssymb, nobibnotes, aps, prl, floatfix]{revtex4-2}
\usepackage{graphicx}

\usepackage{natbib}
\usepackage{bm}          
\usepackage{xcolor}
\usepackage{tikz}
\usepackage{hyperref}
\usepackage{float}
\usepackage{amsmath}
\usepackage{subcaption}

\usepackage[ruled]{algorithm2e}

\usepackage{chngcntr}

\usetikzlibrary{arrows.meta,positioning,decorations.markings}

\begin{document}

\title[LBM for 2D turbulent flows]{Exploring the Lattice-Boltzmann method for two-dimensional turbulence simulation}

\author{Raquel Dapena-García}
\email[Email:~]{raquel.dapena.garcia@usc.es} % optional
\affiliation{CRETUS, Group of Nonlinear Physics, University of Santiago de Compostela, Spain} % optional second address
% If there were a second author at the same address, we would put another 
\author{Vicente Pérez-Muñuzuri} 
% \author statement.
\affiliation{CRETUS, Group of Nonlinear Physics, University of Santiago de Compostela, Spain}

\date{\today}

\begin{abstract}
The Lattice-Boltzmann method is a mesoscopic approach for solving hydrodynamic problems involving laminar and turbulent fluids. Although its suitability for laminar flows is supported by a myriad of studies, turbulent flows give rise to additional challenges. We estimate the accuracy of the simulation results obtained via an implementation of a Lattice-Boltzmann solver for a two-dimensional turbulent flow. We apply this solver to simulate a two-dimensional flow field filled with randomly located rigid disks and study the von Karman vortex street generated after the wake of such obstacles. The implementation of these results is provided as supplementary material.
\end{abstract}

\maketitle

\section{Introduction}

Turbulence plays a crucial role in natural flows. It is ubiquitous in nature, shaping a vast array of physical processes in the atmosphere, oceans, rivers, and in astrophysical environments \citep{Pope2000}. In the atmosphere, turbulent mixing drives the redistribution of energy, moisture, and momentum, significantly influencing weather patterns and climate dynamics \citep{Vallis2017}. Most notably, the chaotic motion of turbulent eddies in clouds enhances the mixing of air masses, leading to cloud formation and precipitation. In rivers and streams, turbulence enhances oxygen transfer and sediment transport, which are vital for aquatic ecosystems and the geomorphological shaping of landscapes \citep{knighton2014fluvial}.

Turbulence is also important in astrophysical and planetary processes. In the formation of stars and galaxies, turbulent gas flows within interstellar clouds control the rate at which matter collapses to form new stars. Similarly, in the oceans \citep{Galperin2010}, turbulence near the surface and at boundaries drives nutrient mixing, supporting marine life by enabling photosynthesis and biological productivity. Turbulence is also important at smaller scales, with turbulent flow patterns observed in larynx and nasal cavities \citep{davidson2015turbulence}, as well as in major arteries such as the aorta, potentially signaling abnormalities in the valve's function \citep{stein1976turbulent}. Although turbulence introduces complexity and unpredictability to fluid systems, its presence is essential for efficient mixing, energy transfer, and the self-organization of natural systems. Without turbulence, many vital processes in nature would be drastically slowed down or rendered ineffective.

The nature of turbulence in three-dimensional (3D) and two-dimensional (2D) flows differs fundamentally in how energy and structures evolve over time \citep{Tabeling2002,Boffetta2012}. In 3D turbulence---as commonly seen in atmospheric flows, oceans, and pipe systems---energy cascades from large scales to smaller ones, a process known as the forward energy cascade. This process leads to the breakdown of large eddies into progressively smaller vortices, where viscous forces eventually dissipate the energy. Thus, energy moves toward the largest wave numbers. This behavior is central to the chaotic and dissipative nature of 3D turbulence and is key to understanding energy loss and mixing in natural and engineered systems \citep{Kolmogorov1941}.

In contrast, 2D turbulence, such as that observed in thin layers of the atmosphere, soap films \citep{Sommeria1982,ParetTabeling1997}, Faraday waves \citep{Kameke2011} and planetary atmospheres such as Jupiter’s, exhibits an inverse energy cascade. In this case, energy injected at small or intermediate scales moves toward larger scales, causing small vortices to merge into larger, more coherent structures. This movement explains the persistence of large-scale features such as Jupiter's Great Red Spot \citep{Marcus1993}. The different behavior arises from constraints on vortex stretching: in 3D, vortices can stretch and twist, enabling energy to move to smaller scales; in 2D, such stretching is absent, which changes the dynamics. These distinctions fundamentally alter how we model and predict fluid behavior in different dimensions.

Turbulent flows are characterized by their energy spectrum $E_k$ and enstrophy spectrum $Z_k$. The former is obtained directly from the Fourier transform of the velocity field $\vec{u}(\vec{r})$, defined as
\begin{equation}
\vec{u}(\vec{k}) = \frac{1}{(2\pi)^{d/2}}
\int_D \vec{u}(\vec{r})\, e^{-i \vec{k}\cdot\vec{r}} \, d\vec{r},
\end{equation}
where $d$ is the spatial dimension. The energy spectrum $E_k$ is defined such that the energy contained in Fourier modes with wave numbers in the interval $[k, k+dk]$ is
\begin{equation}
E_k\, dk = \frac{1}{2} \left| \vec{u}(\vec{k}) \right|^2 \, dk,
\end{equation}
so that the total energy is given by
\begin{equation}
    E=\frac{1}{2}\int_D |\vec{u}|^2 d\vec{r} = \int E_k dk.
    \label{eq:energy}
\end{equation}

Similarly, the enstrophy, a measure of the flow vorticity, is given by
\begin{equation}
    Z=\frac{1}{2}\int_D |\vec{\omega}|^2 = \int k^2 E_k dk = \int Z_k dk,
    \label{eq:enst}
\end{equation}
where $\vec{\omega}=\vec{\nabla}\times \vec{u}$ is the vorticity.

In 3D turbulence standard Kolmogorov scaling applies with $E_k \propto k^\gamma$, $Z_k \propto k^2E_k=k^{\gamma +2}$, and $\gamma=-5/3$ \citep{Kolmogorov1941}. Kraichnan's theory of forced 2D turbulence predicts the development of a double energy cascade \citep{Kraichnan1967}. When the flow is forced at some wavelength $\lambda_f$, the energy is passed on to larger wavelengths generating an inverse energy cascade with the same scaling as in 3D turbulence for $k < k_f$. However, for $k>k_f$ enstrophy is passed to smaller wavelengths and in this range the energy scales as $E_k \propto k^{-3}$ and $Z_k \propto k^{-1}$. In 2D turbulence, energy transfers from small (large $k$)  to large (small $k$) length scales  (opposite to the behavior observed in 3D turbulence), while the enstrophy moves toward higher wave numbers (corresponding to smaller wavelengths); thus, energy and enstrophy propagate in opposite directions through the spectrum.

In computational fluid dynamics \citep{ferziger2002computational,fletcher2012computational}, the simulation of turbulent flows relies on a hierarchy of models that balance accuracy and computational cost. At the highest fidelity, direct numerical simulation \citep{DORAN2013201} of the Navier-Stokes equation resolves all the relevant scales of motion, capturing the full complexity of turbulence without modeling assumptions. However, its computational demands make it impractical for most real-world applications. Alternatively, there are two main families of models, ordered by decreasing complexity and accuracy \citep{wilcox1998turbulence}: Large Eddy Simulation, which solves the full problem from the mean flow down to the larger turbulent eddies (which are the most energetic), while excluding the smallest scales from the numerical simulation and introducing a model to account for the influence of these small scales on the rest of the flow; and the Reynolds-Averaged Navier-Stokes equations, which are based on approximating the apparent Reynolds stresses (or Favre stresses for compressible flows). The choice among these models depends on the desired balance between accuracy and computational cost.

In addition to these traditional approaches, the Lattice-Boltzmann method \citep{succi2001lattice,kruger2017lattice,mohamad2019} has emerged as an alternative practical framework for simulating hydrodynamic problems and has evolved sufficiently to compete with the previously mentioned tools for solving complex fluid flow problems. This method adopts a mesoscopic perspective and models the fluid by particle distribution functions whose collective dynamics approximate macroscopic flow behavior. 

In the following, we simulate a 2D turbulent flow using the Lattice-Boltzmann method to compare the energy and enstrophy Fourier spectra with those predicted by the theory. This comparison will further the understanding of the capabilities of the Lattice-Boltzmann method to treat flows in this regime, while also serving as a gentle introduction to turbulent flows due to the relative straightforward implementation of the algorithm compared to Navier-Stokes solvers. 

\section{The Lattice Boltzmann method for hydrodynamic problems}

Computational fluid dynamics algorithms rely on a numerical discretization of the Navier-Stokes equations, where the quantities of interest are typically the velocity and density of the fluid in each region of the domain. In contrast, the Lattice-Boltzmann method focuses on the collective behavior of the fluid's particles, for which a mesoscopic study of the system is required. Rather than solving the Navier-Stokes equations directly, the Lattice-Boltzmann method is based on the Boltzmann equation, which describes the behavior of a system of particles moving under the influence of forces such as collisions and external fields. The continuous fluid domain is discretized onto a lattice, and the Lattice-Boltzmann equation is solved using a set of discrete velocities \citep{Aidun1998}, which allows for the efficient simulation of complex flows, including turbulent ones. One of the advantages of the Lattice-Boltzmann method is that it can easily handle complex geometries and boundary conditions, and is well-suited for parallelization \citep{kruger2017lattice}.

To illustrate how the Lattice-Boltzmann equation can be derived from the Boltzmann equation, some ideas of kinetic theory are necessary. The kinetic theory of gases is concerned with the average or statistical properties of the particles that form the system. As such, its fundamental variable keeps track of the density of particles moving with a velocity $\bm{\xi}$ within a volume of space $\bm{x}$ at a given time $t$. This variable is called the particle distribution function $f(\bm{x}, \bm{\xi}, t)$, and the equation that governs the evolution of this function is the Boltzmann equation, which can be written in index notation as
\begin{equation}
    \frac{\partial f}{\partial t} + \xi_\alpha \frac{\partial f} {\partial x_\alpha} + \frac{F_\alpha}{\rho} \frac{\partial f}{\partial \xi_\alpha}=\Omega(f),
    \label{eq:be}
\end{equation}
where $\alpha=\{ x,y,z \}$ (repeated indices are summed over), $F$ represents the external force applied to the system, and $\rho$ is the fluid density. Equation~(\ref{eq:be}) states that the redistribution of particles as they move through space (left-hand side) is balanced by their interactions and collisions (right-hand side). Although Eq.~(\ref{eq:be}) might seem deceptively simple, the arbitrary form of the collision operator $\Omega(f)$ makes its solution straightforward for only a specific set of problems, because it takes into account all the possible outcomes of two-particle collisions. As we will explain, because we do not aim to solve the Boltzmann equation directly, we can assume a simplified collision process that greatly simplifies the expression for $\Omega(f)$.

To derive the Lattice-Boltzmann equation from Eq.~(\ref{eq:be}), a numerical discretization method is applied. These methods typically discretize space and time into a lattice with grid spacing $\Delta x$ and time step $\Delta t$. The Lattice-Boltzmann method goes one step further by also discretizing velocity space; that is, it restricts the directions in which the particles can move to a limited set of velocities $\bm{c}_i$. There is a natural trade-off between the number of allowed velocities and the computational cost of the algorithm: choosing more velocities increases accuracy but requires more resources. Therefore, a prior analysis of the type of problem under study must be considered when selecting the velocity set to balance accuracy and efficiency. These approximations are labeled as $DdQq$ models, where $d$ represents the number of spatial dimensions of the system and $q$ the number of velocities.

In this paper we implement a standard force-free Lattice-Boltzmann solver with a D2Q9 (2 spatial dimensions and 9 velocities) lattice as seen in Fig.~\ref{im:d2q9} to describe an incompressible Newtonian fluid. Each velocity $\bm{c}_i$ has an associated weight $w_i$, chosen so that the isotropy of the lattice is preserved. This weighting reflects the intuitive idea that nodes farther from the particle’s current location contribute less to the dynamics, while nearby nodes have a stronger influence. Such a choice helps maintain the correct macroscopic behavior of the fluid.

\begin{figure}[H]
    \centering
    \begin{tikzpicture}[scale=1.2,
        vec/.style={-Stealth, line width=1pt},
        nodepoint/.style={circle, fill=black, inner sep=1.2pt},
        labelsmall/.style={font=\small}
      ]
    
      % Axes
      \draw[dashed,-,opacity=0.5] (-2.2,0) -- (-1.3,0);
      \draw[dashed,->,opacity=0.5] (1.3,0) -- (2.2,0) node[right] {$x$};
      \draw[dashed,-, opacity=0.5] (0,-2.2) -- (0,-1.4);
      \draw[dashed,->, opacity=0.5] (0,1.4) -- (0,2.2) node[above] {$y$};
    
      % Lattice central point
      \coordinate (O) at (0,0);
      \node[nodepoint] at (O) {};
    
      % Unit positions (D2Q9)
      \coordinate (E)  at (1,0);
      \coordinate (W)  at (-1,0);
      \coordinate (N)  at (0,1);
      \coordinate (S)  at (0,-1);
    
      \coordinate (NE) at (1,1);
      \coordinate (NW) at (-1,1);
      \coordinate (SE) at (1,-1);
      \coordinate (SW) at (-1,-1);
    
      % Draw nodes
      \foreach \p in {E,W,N,S,NE,NW,SE,SW} {
        \node[nodepoint] at (\p) {};
      }
    
      % Arrows from origin
      \draw[vec] (O) -- (E)  node[right=1pt, labelsmall] {$6$};
      \draw[vec] (O) -- (W)  node[left=1pt, labelsmall] {$3$};
      \draw[vec] (O) -- (N)  node[above=1pt, labelsmall] {$2$};
      \draw[vec] (O) -- (S)  node[below=1pt, labelsmall] {$1$};
    
      \draw[vec] (O) -- (NE) node[near end, above right=10pt, labelsmall] {$8$};
      \draw[vec] (O) -- (NW) node[near end, above left=10pt,  labelsmall] {$5$};
      \draw[vec] (O) -- (SW) node[near end, below left=10pt,  labelsmall] {$4$};
      \draw[vec] (O) -- (SE) node[near end, below right=10pt, labelsmall] {$7$};
    
      % Draw unit square grid
      \draw[gray!60] (-1,-1) grid (1,1);
    
    \end{tikzpicture}
    \caption{Schematic of the D2Q9 velocity set. A particle located at the central node can move in eight possible directions, either through the diagonal links or the main axis links, or remain stationary at its location. The square marked with gray lines has a length of $2\Delta x$. The rest velocity $0$ located at the central node is not shown. See Table~\ref{tab:vel_set} for more details.}
    \label{im:d2q9}
\end{figure}

\begin{table}[H]
\centering
\begin{tabular}{c|ccccccccc}
\hline
$i$     & 0   & 1   & 2   & 3   & 4   & 5   & 6   & 7   & 8   \\
\hline
$w_i$   & $4/9$ & $1/9$ & $1/9$ & $1/9$ & $1/36$ & $1/36$ & $1/9$ & $1/36$ & $1/36$ \\
$c_{ix}$ & 0 & 0 & 0 & $-1$ & $-1$ & $-1$ & $1$ & $1$ & 1 \\
$c_{iy}$ & 0 & $-1$ & 1 & 0 & $-1$ & 1 & 0 & $-1$ & $1$ \\
\hline
\end{tabular}
\caption{The velocity set $\mathbf{c}_i\ = (c_{ix}, c_{iy})$ and its corresponding weights for the D2Q9 model. All $c_i$ have units of $\Delta x/\Delta t$.}
\label{tab:vel_set}
\end{table}
% and a magnitude of 1, because we set $\Delta x=\Delta t=1$.
Given the discretization of velocity space, the original particle distribution function $f(\bm{x}, \bm{\xi},t)$ becomes $f_i(\bm{x},t)$. To discretize space and time, we apply a forward difference method to Eq.~(\ref{eq:be}). The resulting equation is given by
\begin{equation}
    f_i(\bm{x}+\bm{c}_i\Delta t, t +\Delta t)=f_i(\bm{x},t)-\frac{\Delta t}{\tau}\left(f_i(\bm{x},t)-f_i^{\rm eq}(\bm{x},t)\right),
    \label{eq:lbm}
\end{equation}

\noindent where $f_i(\bm{x},t)$ represents the distribution function of the discrete set of velocities $\bm{c}_i$ at position $\bm{x}$ and time $t$. Here, the fluid dynamics are modeled using the BGK collision operator \citep{PhysRev.94.511}
\begin{equation}
    \Omega_i=-\frac{f_i(\bm{x},t)-f_i^{\rm eq}(\bm{x},t)}{\tau} \Delta t,
    \label{eq:omega_bgk}
\end{equation}
which assumes that the relaxation toward equilibrium can be modeled by a single relaxation time $\tau$.

In Eq.~(\ref{eq:lbm}) the function $f_i^{\rm eq}$ represents the local equilibrium distribution function \citep{Qian_1992} and is described by
\begin{equation}
    f_i^{\rm eq}=\rho w_i \left(1+ \frac{\bm{c}_i\cdot \bm{u}}{c_s^2}+\frac{\left(\bm{c}_i\cdot \bm{u}\right)^2}{2c_s^4}-\frac{\bm{u}\cdot \bm{u}}{2c_s^2}\right),
    \label{eq:fi_eq}
\end{equation}

\noindent where $\bm{u}$ is the mean velocity at the lattice site as computed in Eq.~(\ref{eq:macro_var}) and $c_s$ is the lattice speed of sound, 
\begin{equation}
    c_s=\frac{1}{\sqrt{3}}\frac{\Delta x}{\Delta t}.
\end{equation}

The equilibrium form in Eq.~(\ref{eq:fi_eq}) corresponds to the discretized version of the Maxwell-Boltzmann distribution, with the exponential expanded in a Taylor series and truncated after the first few terms to yield a second-order approximation consistent with the lattice symmetries.

The macroscopic fluid density $\rho$ and velocity $\bm{u}$ can be consistently recovered from the discrete distribution functions through their zeroth and first moments, respectively:
\begin{equation}
    \rho(\bm{x},t) =\sum_{i=0}^{q-1} f_i(\bm{x},t) \quad \rho(\bm{x},t) \bm{u}(\bm{x},t) = \sum_{i=0}^{q-1}  \bm{c}_if_i(\bm{x},t).
    \label{eq:macro_var}
\end{equation}

It can be shown using the Chapman-Enskog analysis that for incompressible or quasi-compressible continuum flows, the Lattice-Boltzmann method recovers the macroscopic behavior described by the Navier-Stokes equations, with the kinematic fluid viscosity $\nu$ directly related to the lattice parameters as \citep{mohamad2019,kruger2017lattice},
\begin{equation}
    \nu^*=c_s^{*2}\left(\tau^*-\frac{1}{2}\right),
    \label{eq:nu}
\end{equation}

\noindent where $*$ represents quantities given in lattice (dimensionless) units. We set $\Delta x^*=\Delta t^*=1$, as is typically done in Lattice-Boltzmann simulations. To avoid cluttering the notation, we will assume that all variables are given in lattice units and therefore omit the symbol $*$.

For turbulent flows, some modifications to the basic Lattice-Boltzmann algorithm have to be made \citep{mohamad2019}. We follow the approach proposed by Smagorinsky \citep{Smagorinsky1963} and introduce the eddy viscosity $\nu_T$ to add to the kinematic viscosity and capture the effect of unresolved turbulent eddies on momentum transfer: 
\begin{equation}
    \nu_T=C_s^2\sqrt{2\sum_{i,j}S_{ij}S_{ij}},
    \label{eq:nu_turb}
\end{equation}

\noindent where
\begin{equation}
    S_{ij}=\frac{1}{2}\left(\frac{\partial u_i}{\partial x_j}+\frac{\partial u_j}{\partial x_i}\right),   
\end{equation}
is the strain-rate tensor and $C_s$ is the Smagorinsky coefficient with $C_s$ between $0.1-0.4$ \citep{mohamad2019}. We have chosen $C_s=0.1$.

We use Eqs.~(\ref{eq:nu}) and (\ref{eq:nu_turb}) to write $\tau$ as
\begin{equation}
    \tau=3(\nu+\nu_T)+0.5.
    \label{eq:tau_turb}
\end{equation}
\noindent The value of $\tau$ now must be calculated at each time step and for each domain cell, leading to an increase in the computational cost.

Figure~\ref{fig:setup} shows the 2D rectangular domain with $H\times L=600\times2000$ cells. A modified Poiseuille-like velocity profile with a small sinusoidal perturbation was imposed along the transverse direction of the flow at the inlet, and a zero-flux boundary condition was considered for the outlet (right boundary), both achieved through the use of Zou/He type boundary conditions at the mesoscopic level \citep{zou1997pressure}. $N$ disks of radius $R$ were located randomly in an intermediate region far enough from the left boundary and the lateral boundaries (solid gray area in Fig.~\ref{fig:setup}). The disks were placed so that there is no overlap between them and there is enough space for the fluid to move in between disks, keeping a minimum of five lattice cells between each disk and between the disks and the top and bottom boundaries. The bounce-back method \citep{succi2001lattice,kruger2017lattice} was implemented on the channel and disk walls to simulate the macroscopic no-slip condition on the fluid. See Appendix~\ref{app:lbm_cc} for a comprehensive description of the boundary conditions mentioned here. The fluid is initialized with a density $\rho=1$, viscosity $\nu=10^{-3}$, and velocity $u_{\max}={\rm Re}\,\nu/R$, with $\rm Re$ the Reynolds number (all given in lattice units). The Reynolds number, which sets the value of $u_{\max}$, is a dimensionless quantity that compares the ratio of inertial forces to viscous forces in a flow, and serves as an indicator of when small perturbations grow instead of being damped and the flow transitions from orderly to chaotic.

\begin{figure}[H]
		
    \centering
    \begin{tikzpicture}[scale=0.65]
    % Parameters
    \def\L{20}
    \def\H{5}
    
    % Channel walls (black)
    \draw[thick] (0,0) -- (\L,0);
    \draw[thick] (0,\H) -- (\L,\H);
    \draw[thick] (0,0) -- (0,\H);
    \draw[thick] (\L,0) -- (\L,\H);
    
    % Inlet arrows
    \foreach \y in {1,2,3,4} {
        \draw[->, thick] (-1,\y) -- (0.5,\y);
    }
    
    % Solid obstacle
    \draw[fill=lightgray] (2.5,0.55) rectangle (8.5,4.45);

    % Region of study
    \draw[fill=lightgray, dashed] (10,0.55) rectangle (19,4.45);
    
    % Particles
    \foreach \pos in {(3.5,1.5),(3.5,2.75),(4.25,2.25)} {
        \draw[thick, fill=white] \pos circle (0.35);
    }
    
    % Dimensions
    \draw[<->] (\L+0.5, 0) -- (\L+0.5, \H);
    \node[right] at (\L+0.5, \H/2) {$H$};
    
    \draw[<->] (0,-0.8) -- (\L,-0.8);
    \node at (\L/2, -1.2) {$L$};
    
    % Boundary labels
    \node[below] at (0,-0.8) {\scriptsize Zou/He (inlet)};
    \node[below] at (\L,-0.8) {\scriptsize Zou/He (outlet)};
    \node[above] at (\L/2,\H) {\scriptsize Bounce-back};
    \node[below] at (\L/2,0) {\scriptsize Bounce-back};
    \node at (6.3,2) {\scriptsize Bounce-back};
    
    \end{tikzpicture}
    \caption{Representation of the simulation setup. $N$ disks are located randomly within the gray area delimited by solid lines (three are shown as an example). No-slip (bounce-back) boundary conditions are imposed at the channel lateral walls and on the surface of the immersed disks. For the inlet and outlet boundaries we enforce velocity and pressure (respectively) boundary conditions using the Zou/He method. The inlet velocity profile is simplified using arrows oriented along the main direction of the flow. The gray region enclosed by dashed lines represents the focus region of this simulation.}
    \label{fig:setup}
\end{figure}

The simulation steps implemented for this study are summarized in Algorithm \ref{alg:lbm_multicyl}. We refer to Appendix~\ref{app:lbm_impl} for additional implementation details. The simulation begins by initializing the Lattice-Boltzmann parameters and randomly positioning non-overlapping cylindrical obstacles (represented by disks in 2D) within the domain. The flow is advanced in time through standard collision and streaming steps with the chosen boundary conditions. At fixed intervals, turbulence-related quantities such as the turbulent kinetic energy, vorticity, and energy spectra are computed and stored for analysis. The main region of study is highlighted by a dashed gray rectangle in Fig.~\ref{fig:setup} and begins at the channel center point.

\begin{algorithm}[H]
\caption{Lattice-Boltzmann simulation with multiple cylindrical obstacles}
\label{alg:lbm_multicyl}
\SetAlgoLined
\SetKwInOut{Input}{Input}
\SetKwInOut{Output}{Output}
\SetInd{0.4em}{1em}
\DontPrintSemicolon

\Input{Number of cylinders $N_\text{cyl}$ and their radius $r$}
\Output{Time evolution of turbulence parameters, energy spectra, and flow visualizations}

\BlankLine
\textbf{Setup:} \\
Set simulation parameters: system size $(n_x, n_y)$, number of iterations \texttt{maxIter}, \texttt{Re} number and viscosity $\nu$\;
Place disk at random while avoiding overlap and define a boolean array on the lattice that identifies which nodes are obstacles (inside a disk) or fluid (outside).\;

\BlankLine
\textbf{Initialization:} \\
Initialize the velocity field $\bm{u}(x,y)$ profile\;
Compute initial equilibrium distributions $f_i^{\rm eq}$ and set $f_i = f_i^{\rm eq}$\;

\BlankLine
\textbf{Main time loop:} \\
\For{$t = 0$ \KwTo $\texttt{maxIter}$}{
  Compute macroscopic variables $\rho$ and $\mathbf{u}$ from $f_i$ via Eq.~(\ref{eq:macro_var})\;
  Compute $f_i^{\rm eq}$ using Eq.~(\ref{eq:fi_eq})\;
  Compute strain tensor $S_{ij}$ and eddy viscosity $\nu_T$\;
  Perform collision step: find post-collision populations $f_i^*$ following Eq.~(\ref{eq:collision})\;
  Apply bounce-back boundary condition on obstacles\;
  Perform streaming step: shift $f_i^*$ along $\mathbf{c}_i$ according to Eq.~(\ref{eq:streaming})\;
  Apply boundary conditions as needed and recalculate the populations at the inlet and outlet\;
  If desired, compute and store flow statistics (e.g., turbulent kinetic energy, energy spectra)\;
}

\BlankLine
\textbf{End simulation}

\end{algorithm}

\newpage
    
\section{Results}
\label{app:lbm_results}

\begin{figure}[H]
\centering
\includegraphics[width=0.8\linewidth]{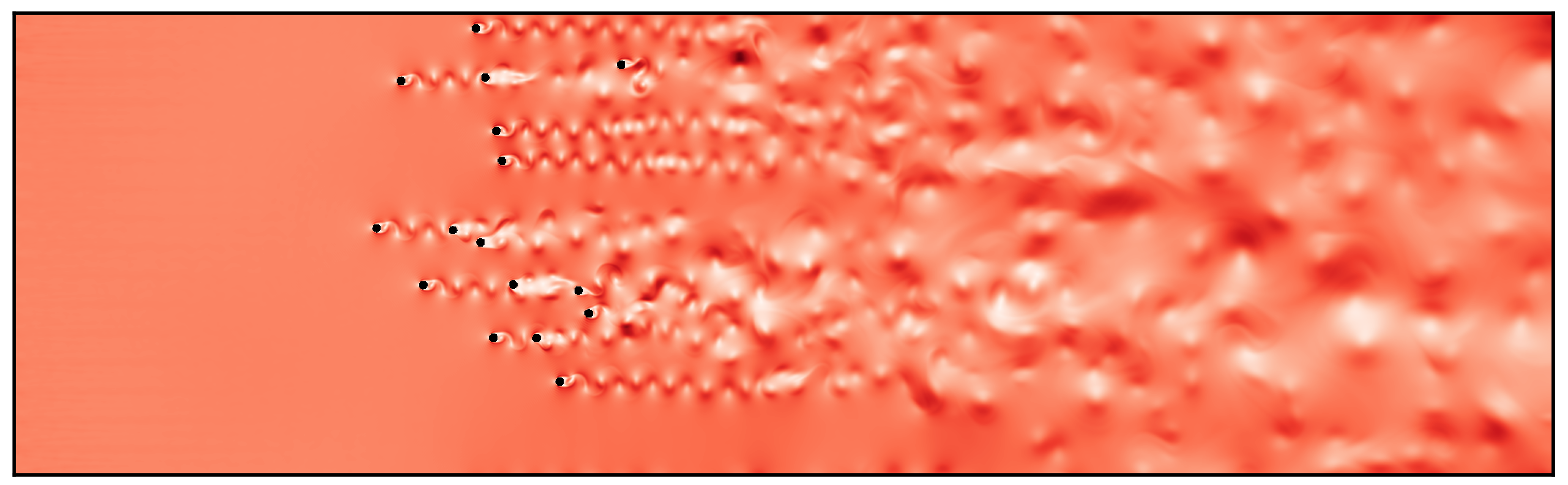} \\
\includegraphics[width=0.8\linewidth]{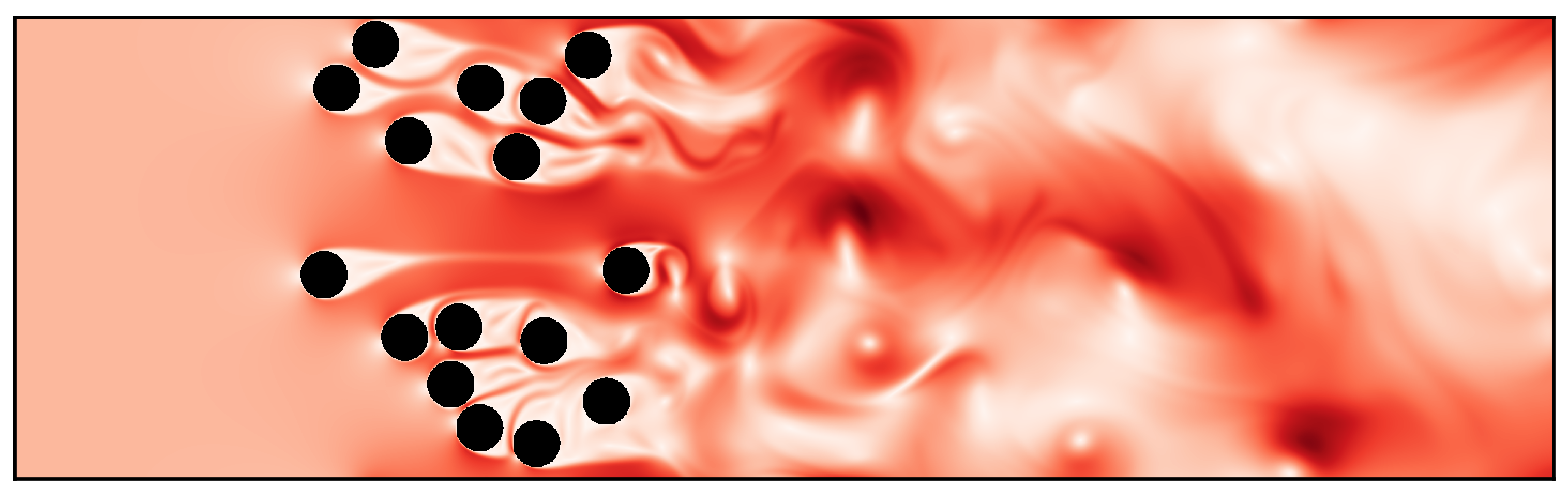} \\
\includegraphics[width=0.8\linewidth]{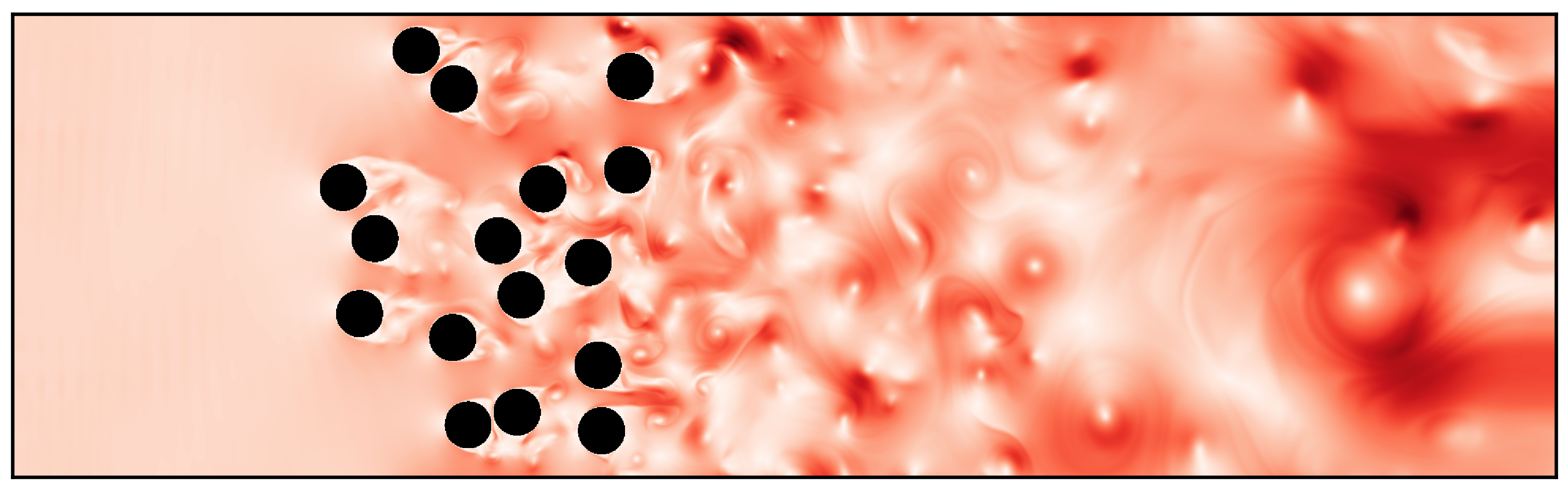} 
\caption{Visualization of a 2D flow at a given time passing through an array of disks for different disk sizes and Reynolds number: $R=5$, $\rm Re=200$ (top); $R=30$, $\rm Re=200$ (middle); and $R=30$, $\rm Re=800$ (bottom). Number of disks $N=16$ for all three examples.}
\label{fig:vortices}
\end{figure}

The vortical structures that develop in the wake of the disks are shown in Fig.~\ref{fig:vortices}. Note that in the vicinity of the disks, several von Karman vortex streets \citep{Pope2000,kundu2024fluid} (a repeating pattern of swirling vortices) develop and interact with each other, while further downstream these vortices start to blend together into bigger swirling patterns. As expected, the size of the vortices increases with the disk radius, with larger disks producing larger vortices. An increase in the Reynolds number leads to higher velocities and stronger vortex interactions in the array wake. This behavior, where small structures merge into larger ones, is characteristic of turbulence in 2D flows.

\begin{figure}[H]
\centering
\includegraphics[width=\linewidth]{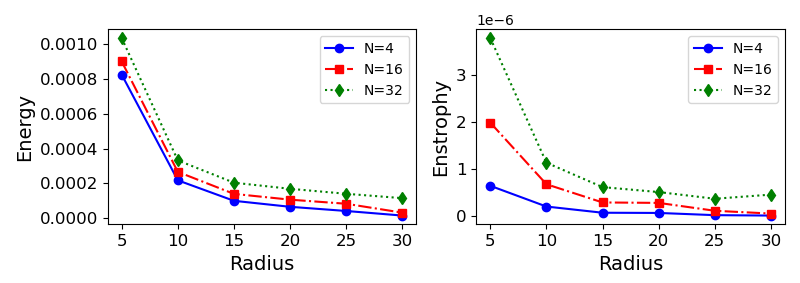}
\caption{Mean spatio-temporal kinetic energy and enstrophy values as a function of the disk radius for different numbers of disks. Reynolds number $\rm Re=200$.}
\label{fig:EZ_radio}
\end{figure}

Spatio-temporal energy and enstrophy averages calculated downstream of the disks are shown in Fig.~(\ref{fig:EZ_radio}) as a function of the radius of the disks. In these simulations, $u_{\max}$ decreases as $R$ increases, because the Reynolds number and viscosity remain constant. For this reason, to compare results for the time averages, we choose the same time duration after the vortex street is fully developed in the domain. 

Both the energy and enstrophy decrease with increasing disk size, but increase with the number of disks. Increasing the radius of the disks (while keeping the number fixed) leads to a blockage effect, thus reducing the available space for the fluid to accelerate. This causes a reduction in the velocity fluctuations and, consequently, leads to lower kinetic energy. Additionally, the blockage suppresses vortex shedding, resulting in a net decrease in small-scale vorticity. In contrast, increasing the number of disks (while keeping the radius fixed) generates more vortex wakes that interact, therefore facilitating fine-scale mixing and vortex merging, which in turn increases both the kinetic energy and the enstrophy. These differences are larger for smaller disk radii.

\begin{figure}[H]
\centering
\includegraphics[width=\linewidth]{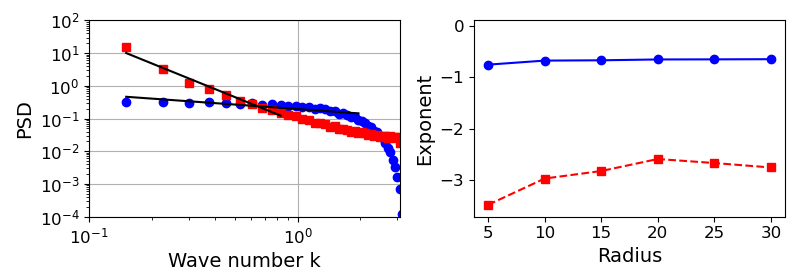}
\caption{(a) Power spectral density (PSD) for the kinetic energy and enstrophy (b) The scaling exponents $\gamma$ for $E_k$ and $Z_k$. The blue solid lines and dots correspond to the enstrophy and the red dashed lines and squares to the energy, respectively. Black lines in (a) correspond to $\log E_k$ ($\log Z_k$) versus $\log k$. Parameters: $\rm Re=200$, $R=20$ and $N=16$.}
\label{fig:PSD_radio}
\end{figure}

Figure~\ref{fig:PSD_radio} shows the energy and enstrophy spectra and their corresponding slope $\gamma$ for different disk radii. This power spectral density describes how the energy of a signal or field is distributed over wave numbers or frequencies. 
For the velocity field $\vec{u}(\vec{r})$, the power spectral density (PSD) is directly related to its Fourier transform $\vec{u}(\vec{k})$, so that $\text{PSD}(k) \sim |\vec{u}(\vec{k})|^2$. The slope for the energy is much steeper and is close to $\gamma\approx -3$, but the slope for the enstrophy is shallower, $\gamma \approx -0.8$, exceeding the theoretical value of $\gamma \approx -1$. Deviations from Kraichnan scaling in our results are consistent with previous numerical studies \citep{lilly1972numerical,herring1974decay} of 2D turbulence, which often report spectral slopes steeper than the theory suggests. The steeper slope may be caused by the appearance of long lived, isolated, axisymmetric vortices \citep{Mcwilliams_1984}.

\begin{figure}[H]
\centering
\includegraphics[width=\linewidth]{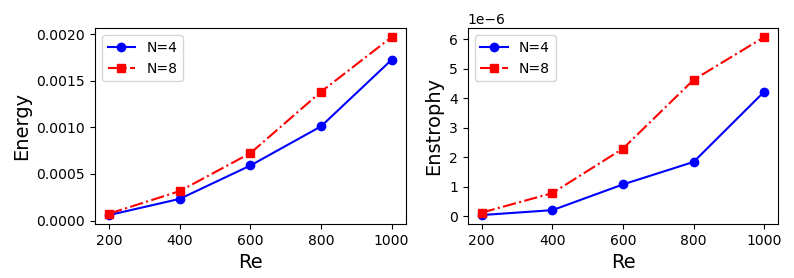}
\caption{(a) Mean values of the spatio-temporal kinetic energy and (b) enstrophy as a function of the Reynolds number for various numbers of disks. Disk radius $R=20$.}
\label{fig:EZ_Reynolds}
\end{figure}

The effect of increasing the Reynolds number on the kinetic energy and enstrophy is shown in Fig.~\ref{fig:EZ_Reynolds}. As expected, both the kinetic energy and enstrophy increase with the Reynolds number, as well as with the number of disks. The values of the exponent $\gamma$ for both variables do not change with $\rm Re$ and their values are qualitatively the same as those shown in Fig.~\ref{fig:PSD_radio}.

\section{Conclusions}

We have explored the capability of the Lattice-Boltzmann method to simulate a two-dimensional turbulent flow that develops in the wake of a set of disks. The exponents for the energy and enstrophy spectra were computed for different Reynolds numbers and disks of varying initial configurations, yielding values of $\gamma\approx -3$ and $\gamma\approx -1$, respectively, close to those predicted by the theory. 

Because the front velocity increases with the Reynolds number, we considered an eddy viscosity function of the strain tensor to better account for the turbulent effects. The results obtained for the exponents show no qualitative differences using a simplified model with a constant value for $\tau$.

Despite capturing the main qualitative features of 2D turbulence, our results show small quantitative deviations from the theoretical predictions for the energy and enstrophy scaling exponents. These discrepancies are primarily attributed to limitations of our numerical implementation. Achieving closer agreement with the theory would require a more sophisticated treatment of the boundary conditions, particularly to ensure an accurate resolution of the flow in the vicinity of the disk-shaped obstacles and near the corners of the computational domain. Although the boundary conditions employed in this work—--namely bounce-back and Zou/He—--are appropriate and widely used, reproducing the theoretical predictions more quantitatively would necessitate their more advanced formulations, which better capture geometric details and near-boundary flow effects. Such deviations are also consistent with earlier numerical studies of 2D turbulence, which caution that apparent inertial-range power laws observed at finite Reynolds numbers may not reflect the true asymptotic behavior. Accordingly, although the simulations reproduce the correct cascade phenomenology and large-scale dynamics, the measured scaling exponents should be interpreted as effective, rather than asymptotic, quantities, reflecting a compromise between computational simplicity and physical fidelity.

The simplified model we have discussed allows students to gain a deeper understanding of the underlying concepts and governing equations of two-dimensional turbulent flows. To support this learning process, we have provided the necessary code to reproduce the results we have discussed. Python was chosen due to its accessibility and wide adoption.

\section{Suggested problems}

\begin{enumerate}
  \item Derive the Lattice-Boltzmann equation, Eq.~(\ref{eq:lbm}), directly from the force-free continuous Boltzmann equation given by Eq.~(\ref{eq:be}) to gain a deeper understanding of the assumptions and approximations underlying the discrete formulation. Hint: Expand $f_i(\mathbf{x} + \mathbf{c}_i \Delta t, t + \Delta t)$ around $(\mathbf{x}, t)$ to approximate derivatives using a first order Taylor expansion. Consider the general expression for the collision term $\Omega_i$ and discuss how first-order and second-order discretizations of this term affect the resulting Lattice-Boltzmann equation.
  
  \item We performed simulations for a range of Reynolds numbers to study how the characteristic flow exponents change as $\rm Re$ increases. Repeat this analysis keeping the velocity constant instead of the kinematic viscosity. You should find that, although $E$ and $Z$ exhibit the opposite trend from Fig.~\ref{fig:EZ_radio} (increasing as the $R$ increases), the overall exponents for both quantities remain the same as those presented in Fig.~\ref{fig:PSD_radio}(b).

  \item Place a single disk in the channel and calculate the frequency of vortex shedding as quantified by the Strouhal number $\rm Sr$,
  \begin{equation}
      {\rm Sr}=\frac{fR}{u},
  \end{equation}
  which relates the vortex shedding frequency $f$ to the characteristic particle radius $R$ and flow velocity $u$. Check that it increases linearly with the disk's Reynolds number $\rm Re_p=uR/\nu$ (in the moderate Reynolds number regime). Hint: To measure the shedding frequency, select a point downstream of the obstacle, record the time series of the local flow velocity, and then calculate its frequency of oscillations using the previously described power spectral density analysis.
  
  \item Simulate alternative obstacle geometries, such as rectangular shapes with different profile inclinations, to analyze the influence of the morphology of the obstacles on the resulting flow dynamics and wake structures. To do so, modify the Setup step from Algorithm~\ref{alg:lbm_multicyl} to place these new obstacles instead of the disks and replicate the figures above.
\end{enumerate}

\appendix*
\setcounter{equation}{0}
\section*{Appendix: LBM Implementation Details} \label{app:lbm_impl}

The Lattice-Boltzmann equation, Eq.~(\ref{eq:lbm}), can be conveniently expressed as a two-step process involving collision and streaming operations. In the collision step, the distribution functions at each lattice node relax toward their equilibrium values according to
\begin{equation}
    f_i^*(\bm{x},t)=f_i(\bm{x},t)-\frac{\Delta t}{\tau}\left(f_i(\bm{x},t)-f_i^{\rm eq}(\bm{x},t)\right).
    \label{eq:collision}
\end{equation}
This step is local, because it depends solely on information available at a single lattice node.

Following the collision step, the streaming step advects the particles to the next node along the lattice links,
\begin{equation}
    f_i(\bm{x}+\bm{c}_i\Delta t, t +\Delta t)=f_i^*(\bm{x},t),
    \label{eq:streaming}
\end{equation}
thereby redistributing information between neighboring nodes.

The decomposition of the algorithm into these two steps, collision and streaming, results in a clear separation between a local process, Eq.~(\ref{eq:collision}), and a linear propagation step, Eq.~(\ref{eq:streaming}). This feature is one of the main strengths of the Lattice-Boltzmann method, because it enables a straightforward implementation on parallel computing architectures. The collision step can be computed independently at each lattice node, while the streaming step involves only direct memory transfers between neighboring nodes. This high degree of spatial locality, combined with the simplicity of data exchange patterns, allows the Lattice-Boltzmann method to achieve near-ideal scalability on both shared-memory and distributed-memory systems. Consequently, the method is particularly well-suited for large-scale simulations of complex fluid flows, for which computational efficiency and parallel performance are essential.

\section*{Appendix: Boundary Conditions in Lattice-Boltzmann simulations} \label{app:lbm_cc}

\subsection*{Bounce-Back boundary conditions}
In fluid dynamics, the boundary conditions typically state the value of the macroscopic density and/or velocity at the domain boundaries. We know from Eq.~(\ref{eq:macro_var}) how to compute those macroscopic quantities from the discrete populations $f_i$. However, the opposite procedure is not unique, and there are many ways that these quantities could be computed. This non-uniqueness property is the reason why so many implementations of boundary conditions are available for Lattice-Boltzmann methods.

When modeling solid stationary or moving boundaries, the bounce-back boundary condition is the most common choice due to its simplicity of implementation. The core idea is that all populations hitting a rigid boundary during the streaming step are simply reflected back to where they came from, thus being reintroduced into the bulk of the domain again (see Fig.~\ref{fig:bb}). 

\begin{figure}[H]
\centering
\begin{subfigure}{0.45\textwidth}
\begin{tikzpicture}[scale=1.2,>=stealth]

% Grey solid region
\fill[gray!30] (-2,-1) rectangle (2,0);

% Dashed wall line
\draw[thick] (-2,0) -- (2,0);

% Top fluid node (bounce-back location)
\draw[] (0,0.6) circle (2pt);

% Other fluid nodes (top row)
\draw ( -1.2,0.6) circle (2pt);
\draw ( -0.6,0.6) circle (2pt);
\draw (  1.2,0.6) circle (2pt);
\draw ( 0.6,0.6) circle (2pt);

% Solid nodes (filled black)
\foreach \x in {-1.2, -0.6, 0,0.6,  1.2}{
    \fill (\x,0) circle (2.5pt);
}

% Incoming velocity arrows
\draw[->,thick] (0,0.6) -- (0,1.17) node[pos=0.85, above=4pt] {$\bm{c}_2$};
\draw[->,thick] (0,0.6) -- (-0.57,1.17) node[pos=0.85, above left=4pt]{$\bm{c}_5$};
\draw[->,thick] (0,0.6) -- (0.57,1.17) node[pos=0.85, above right=4pt]{$\bm{c}_8$};
\draw[->,thick] (0,0.6) -- (0.57,0.6) node[pos=0.85, right=4pt] {$\bm{c}_6$};
\draw[->,thick] (0,0.6) -- (-0.57,0.6) node[pos=0.85, left=4pt]{$\bm{c}_3$};
\draw[->,thick] (0,0.6) -- (0,0.03) node[pos=0.85, below=4pt]{$\bm{c}_1$};
\draw[->,thick] (0,0.6) -- (-0.57,0.03) node[pos=0.85, below left=4pt]{$\bm{c}_4$};
\draw[->,thick] (0,0.6) -- (0.57,0.03) node[pos=0.85, below right=4pt]{$\bm{c}_7$};

% Labels x_b
\node[left] at (-2,0) {$\bm{x}_b$};

% Distribution function label
% \node at (0,-1.3) {$f_i^\star(\bm{x}_b,t)$};

\node[right,font=\scriptsize] at (-2,-0.3) {Wall};
\node[right,font=\scriptsize] at (-2,0.3) {Fluid};

\end{tikzpicture}
% \caption{Incoming populations toward the wall}
\end{subfigure}
\hfill
\begin{subfigure}{0.45\textwidth}
\begin{tikzpicture}[scale=1.2,>=stealth]

% Grey solid region
\fill[gray!30] (-2,-1) rectangle (2,0);

% Dashed wall line
\draw[thick] (-2,0) -- (2,0);

% Top fluid node (bounce-back location)
\draw[] (0,0.6) circle (2pt);

% Other fluid nodes (top row)
\draw ( -1.2,0.6) circle (2pt);
\draw ( -0.6,0.6) circle (2pt);
\draw (  1.2,0.6) circle (2pt);
\draw ( 0.6,0.6) circle (2pt);

% Solid nodes (filled black)
\foreach \x in {-1.2, -0.6, 0,0.6,  1.2}{
    \fill (\x,0) circle (2.5pt);
}

% Incoming velocity arrows
\draw[->,thick] (0,0.6) -- (0,1.17) node[pos=0.85, above=4pt] {$\bm{c}_2$};
\draw[->,thick] (0,0.6) -- (-0.57,1.17) node[pos=0.85, above left=4pt]{$\bm{c}_5$};
\draw[->,thick] (0,0.6) -- (0.57,1.17) node[pos=0.85, above right=4pt]{$\bm{c}_8$};
\draw[->,thick] (0,0.6) -- (0.57,0.6) node[pos=0.85, right=4pt] {$\bm{c}_6$};
\draw[->,thick] (0,0.6) -- (-0.57,0.6) node[pos=0.85, left=4pt]{$\bm{c}_3$};
\draw[->,thick,blue] (0,0.03) -- (0,0.6) node[pos=0.15, below=4pt]{$\text{-}\bm{c}_1$};
\draw[->,thick,blue] (-0.57,0.03) -- (0,0.6) node[pos=0.15, below left=4pt]{$\text{-}\bm{c}_4$};
\draw[->,thick,blue] (0.57,0.03) -- (0,0.6) node[pos=0.15
65, below right=4pt]{$\text{-}\bm{c}_7$};

% Labels x_b
\node[left] at (-2,0) {$\bm{x}_b$};

% Distribution function label
% \node at (0,-1.3) {$f_i^\star(\bm{x}_b,t)$};

\node[right,font=\scriptsize] at (-2,-0.3) {Wall};
\node[right,font=\scriptsize] at (-2,0.3) {Fluid};

\end{tikzpicture}
\end{subfigure}
\caption{Sketch of the bounce-back boundary condition at a bottom wall for a D2Q9 lattice. The populations hitting a wall at a certain time step (left) are reflected back into the domain with opposite directions (blue arrows on right). The gray-shaded area represents the solid region, and the straight line full of boundary nodes $\bm{x_b}$ separates the solid region from the fluid nodes.}
\label{fig:bb}
\end{figure}

Figure ~\ref{fig:bb} illustrates how this method uses the information already known by the system to calculate the incoming populations coming from outside the system. Once the outgoing populations $(f_4,f_1,f_7)$ reach the solid wall, these populations are reflected back with a velocity $\bm{c}_{\bar{i}}=-\bm{c}_i$ as $(f_5,f_2,f_8)$ toward the nearest inside node. The standard collision step Eq.~(\ref{eq:collision}) is replaced by

\begin{equation}
    f_i^*(\bm{x}_b,t)=f_{\bar{i}}(\bm{x}_b,t).
    \label{eq:bb}
\end{equation}

\subsection{Zou/He boundary conditions}

In the Lattice-Boltzmann method, the particle distribution functions are decomposed into equilibrium and non-equilibrium components,
\begin{equation}
f_i = f_i^{\rm eq} + f_i^{\rm neq},
\qquad 
f_i^{\rm neq} \equiv f_i - f_i^{\rm eq},
\label{eq:fi_neq}
\end{equation}
where $f_i^{\rm eq}$ is the local Maxwell-Boltzmann equilibrium given by Eq.~(\ref{eq:fi_eq}).  
Zou and He \citep{zou1997pressure} showed that velocity- and pressure-type boundary conditions can be enforced by prescribing the macroscopic quantities $(\rho,\mathbf{u})$ and imposing bounce-back on the non-equilibrium populations for all unknown incoming directions. Figure~\ref{im:zou_he} illustrates the Zou/He velocity and pressure boundary conditions for a D2Q9 lattice.

\begin{figure}[H]
\centering
\begin{tikzpicture}[scale=1.2,>=stealth]

%----------------------------------------
% Parameters
%----------------------------------------
\def\numnodes{7}             % total inlet/outlet nodes
\def\topY{1.65}              % top wall y
\def\botY{-1.65}             % bottom wall y
\pgfmathsetmacro{\nodespacing}{(\topY-\botY)/(\numnodes-1)}  % uniform spacing
\pgfmathsetmacro{\arrowL}{0.9 * \nodespacing}                % arrow length proportional to spacing

%----------------------------------------
% WALLS + GREY REGIONS
%----------------------------------------
\draw[thick] (0,\topY) -- (7,\topY);
\draw[thick] (0,\botY) -- (7,\botY);

\fill[gray!30] (0,\topY) rectangle (7,\topY+0.6);
\fill[gray!30] (0,\botY) rectangle (7,\botY-0.6);

%----------------------------------------
% INLET nodes (left), 7 evenly spaced
%----------------------------------------
\foreach \i in {0,...,\numexpr\numnodes-1}{
    \pgfmathsetmacro{\y}{\topY - \i * \nodespacing}
    \filldraw (0,\y) circle (2pt);
}

% Inlet label BELOW the array
\node[anchor=north] at (-0.5,-1.65) {Inlet};

%----------------------------------------
% OUTLET nodes (right), 7 evenly spaced
%----------------------------------------
\foreach \i in {0,...,\numexpr\numnodes-1}{
    \pgfmathsetmacro{\y}{\topY - \i * \nodespacing}
    \filldraw (7,\y) circle (2pt);
}

% Outlet label BELOW the array
\node[anchor=north] at (7.5,-1.65) {Outlet};

%----------------------------------------
% D2Q9 at middle inlet node
%----------------------------------------
\pgfmathsetmacro{\yC}{\topY - 3 * \nodespacing} %index node is 3
\coordinate (C) at (0,\yC);

% Cardinal directions
\draw[->,dashed] (C) -- ++(\arrowL,0) node[right] {$f_6$};
\draw[->,thick] (C) -- ++(-\arrowL,0) node[left] {$f_3$};
\draw[->,thick] (C) -- ++(0,\arrowL) node[above] {$f_2$};
\draw[->,thick] (C) -- ++(0,-\arrowL) node[below] {$f_1$};

% Diagonals
\draw[->,dashed] (C) -- ++(\arrowL,\arrowL) node[above right] {$f_8$};
\draw[->,thick] (C) -- ++(-\arrowL,\arrowL) node[above left] {$f_5$};
\draw[->,thick] (C) -- ++(-\arrowL,-\arrowL) node[below left] {$f_4$};
\draw[->,dashed] (C) -- ++(\arrowL,-\arrowL) node[below right] {$f_7$};

%----------------------------------------
% D2Q9 at middle outlet node
%----------------------------------------
\coordinate (Cout) at (7,\yC);

% Cardinal directions
\draw[->,thick] (Cout) -- ++(\arrowL,0) node[right] {$f_6$};
\draw[->,dashed] (Cout) -- ++(-\arrowL,0) node[left] {$f_3$};
\draw[->,thick] (Cout) -- ++(0,\arrowL) node[above] {$f_2$};
\draw[->,thick] (Cout) -- ++(0,-\arrowL) node[below] {$f_1$};

% Diagonals
\draw[->,thick] (Cout) -- ++(\arrowL,\arrowL) node[above right] {$f_8$};
\draw[->,dashed] (Cout) -- ++(-\arrowL,\arrowL) node[above left] {$f_5$};
\draw[->,dashed] (Cout) -- ++(-\arrowL,-\arrowL) node[below left] {$f_4$};
\draw[->,thick] (Cout) -- ++(\arrowL,-\arrowL) node[below right] {$f_7$};

\end{tikzpicture}

\caption{Schematic illustration of the Zou/He boundary conditions for a D2Q9 lattice.
Left: inlet boundary where the macroscopic velocity is prescribed.
Right: outlet boundary where the macroscopic pressure (or density) is prescribed.
Solid arrows indicate known particle distribution functions after streaming, while dashed arrows denote unknown incoming populations reconstructed using the Zou/He approach.}
\label{im:zou_he}
\end{figure}

% \subsubsection*{Non-Equilibrium Bounce-Back Rule}

We will denote $\bar{i}$ as the direction opposite to $i$.  
For each unknown boundary population $f_i$ (incoming to the domain), Zou/He applies the non-equilibrium bounce-back rule
\begin{equation}
f_i^{\rm neq} = f_{\bar{i}}^{\rm neq}
\qquad\Longleftrightarrow\qquad
f_i = f_i^{\rm eq} + \big( f_{\bar{i}} - f_{\bar{i}}^{\rm eq} \big).
\label{eq:fi_neq2}
\end{equation}
Equation~(\ref{eq:fi_neq2}) is the compact form of the Zou/He boundary condition. All explicit formulas for specific boundaries follow from this identity once the boundary macroscopic quantities are known.

% \subsubsection*{Velocity Inlet (Left Boundary)}
At a left velocity boundary, the incoming unknown populations are $(f_8,\; f_6,\; f_7)$. The prescribed macroscopic velocities are
\begin{equation}
u_x = u_x^{\text{in}},\qquad u_y = u_y^{\text{in}}. 
\end{equation}

By using the definitions of the macroscopic density and momentum in Eq.~(\ref{eq:macro_var}), we can solve for the unknown populations
\begin{subequations}
\begin{align}
% $$\rho=f_0+f_1+f_2+f_3+f_4+f_5+f_6+f_7+f_8$$
f_6+f_7+f_8&=\rho-(f_0+f_1+f_2+f_3+f_4),\\
% $$\rho u_x=f_6-f_3+f_8+f_7-f_5-f_4$$
f_6+f_7+f_8&=\rho u_x+(f_3+f_5+f_4),\\
% $$\rho u_y=f_2-f_1+f_8-f_7+f_5-f_4$$
f_8-f_7&=\rho u_y-f_2+f_1-f_5+f_4,
\end{align}
\label{eq:system_inlet}
\end{subequations}
using the information already known to the system (the right-hand side of Eq.~(\ref{eq:system_inlet})).

After streaming, the density is obtained from mass conservation:
\begin{equation}
\rho = \frac{f_0 + f_1 + f_2 +f_5 + 2(f_3 + f_4)}{
1 - u_x^{\text{in}}}.
\end{equation}
However, $(f_8,\; f_6,\; f_7)$ remain undetermined. To close the system, we assume the bounce-back rule is still correct for the non-equilibrium part of the particle distribution normal to the boundary. Using Eq.~(\ref{eq:fi_neq}) this assumption implies that
\begin{equation}
    f_6-f_6^{\rm eq}=f_3-f_3^{\rm eq},
\label{eq:fi_normal}
\end{equation}
where $f_3$ is known from streaming.

Once the macroscopic density $\rho$ and velocities $(u_x^{\rm in}, u_y^{\rm in})$ are specified, the equilibrium distributions $f_i^{\rm eq}$ are computed and the unknown populations are reconstructed using the non-equilibrium bounce-back rule:
\begin{subequations}
\begin{align}
f_8 &= f_8^{\rm eq} + \big(f_5 - f_5^{\rm eq}\big), \\
f_6 &= f_6^{\rm eq} + \big(f_3 - f_3^{\rm eq}\big), \\
f_7 &= f_7^{\rm eq} + \big(f_4 - f_4^{\rm eq}\big).
\end{align}
\end{subequations}

% \subsubsection*{Pressure Outlet (Right Boundary)}

At a right pressure boundary, the incoming unknown populations are $(f_5, f_3, f_4)$. The prescribed macroscopic quantity is the density (pressure):
\begin{equation}
    \rho = \rho^{\rm in}=\rho^{\rm out}.
\end{equation}

After streaming, the unknown populations satisfy:
\begin{subequations}
\begin{align}
f_5+f_3+f_4&=\rho-(f_0+f_1+f_2+f_6+f_7+f_8),\\
u_x&=1-\frac{f_7+f_1+f_2+2(f_0+f_8+f_6)}{\rho},\\
u_y&=\frac{f_0+f_2+f_8+(f_5-f_4)-f_1-f_7}{\rho}.
\end{align}
\end{subequations}

Once the macroscopic velocity $(u_x, u_y)$ is reconstructed, the equilibrium distributions $f_i^{\rm eq}$ are computed.  The unknown incoming populations are then determined using the non-equilibrium bounce-back rule along each normal direction [Eq.~(\ref{eq:fi_normal})], e.g., with $f_6, f_7, f_8$ known from streaming:
\begin{subequations}
\begin{align}
f_5 &= f_5^{\rm eq} + \big(f_8 - f_8^{\rm eq}\big), \\
f_3 &= f_3^{\rm eq} + \big(f_6 - f_6^{\rm eq}\big), \\
f_4 &= f_4^{\rm eq} + \big(f_7 - f_7^{\rm eq}\big).
\end{align}
\end{subequations}

\section*{Data availability}
The code with the Lattice-Boltzmann method algorithm can be found in \url{https://github.com/NonlinPhysGroup/Turbulence2D} and in the supplementary material.

\section*{Author declarations}
\noindent\textbf{Conflict of Interests:} The authors have no conflicts to disclose.

\noindent\textbf{Author Contributions:} All authors contributed to writing the original manuscript.

\input{Turb2D_LB_vf_arXiv.bbl}
%\bibliography{LBM}

\end{document}

%% file: Turb2D_LB_vf_arXiv.bbl

%% file: Turb2D_LB_vf_arXiv.bbl
\begin{thebibliography}{30}%
\makeatletter
\providecommand \@ifxundefined [1]{%
 \@ifx{#1\undefined}
}%
\providecommand \@ifnum [1]{%
 \ifnum #1\expandafter \@firstoftwo
 \else \expandafter \@secondoftwo
 \fi
}%
\providecommand \@ifx [1]{%
 \ifx #1\expandafter \@firstoftwo
 \else \expandafter \@secondoftwo
 \fi
}%
\providecommand \natexlab [1]{#1}%
\providecommand \enquote  [1]{``#1''}%
\providecommand \bibnamefont  [1]{#1}%
\providecommand \bibfnamefont [1]{#1}%
\providecommand \citenamefont [1]{#1}%
\providecommand \href@noop [0]{\@secondoftwo}%
\providecommand \href [0]{\begingroup \@sanitize@url \@href}%
\providecommand \@href[1]{\@@startlink{#1}\@@href}%
\providecommand \@@href[1]{\endgroup#1\@@endlink}%
\providecommand \@sanitize@url [0]{\catcode `\\12\catcode `\$12\catcode
  `\&12\catcode `\#12\catcode `\^12\catcode `\_12\catcode `\%12\relax}%
\providecommand \@@startlink[1]{}%
\providecommand \@@endlink[0]{}%
\providecommand \url  [0]{\begingroup\@sanitize@url \@url }%
\providecommand \@url [1]{\endgroup\@href {#1}{\urlprefix }}%
\providecommand \urlprefix  [0]{URL }%
\providecommand \Eprint [0]{\href }%
\providecommand \doibase [0]{https://doi.org/}%
\providecommand \selectlanguage [0]{\@gobble}%
\providecommand \bibinfo  [0]{\@secondoftwo}%
\providecommand \bibfield  [0]{\@secondoftwo}%
\providecommand \translation [1]{[#1]}%
\providecommand \BibitemOpen [0]{}%
\providecommand \bibitemStop [0]{}%
\providecommand \bibitemNoStop [0]{.\EOS\space}%
\providecommand \EOS [0]{\spacefactor3000\relax}%
\providecommand \BibitemShut  [1]{\csname bibitem#1\endcsname}%
\let\auto@bib@innerbib\@empty
%</preamble>
\bibitem [{\citenamefont {Pope}(2000)}]{Pope2000}%
  \BibitemOpen
  \bibfield  {author} {\bibinfo {author} {\bibfnamefont {S.}~\bibnamefont
  {Pope}},\ }\href@noop {} {\emph {\bibinfo {title} {Turbulent Flows}}}\
  (\bibinfo  {publisher} {Cambridge University Press},\ \bibinfo {address}
  {London},\ \bibinfo {year} {2000})\BibitemShut {NoStop}%
\bibitem [{\citenamefont {Vallis}(2017)}]{Vallis2017}%
  \BibitemOpen
  \bibfield  {author} {\bibinfo {author} {\bibfnamefont {G.}~\bibnamefont
  {Vallis}},\ }\href@noop {} {\emph {\bibinfo {title} {Atmospheric and Oceanic
  Fluid Dynamics. Fundamentals and large-scale circulation}}}\ (\bibinfo
  {publisher} {Cambridge University Press},\ \bibinfo {address} {London},\
  \bibinfo {year} {2017})\BibitemShut {NoStop}%
\bibitem [{\citenamefont {Knighton}(2014)}]{knighton2014fluvial}%
  \BibitemOpen
  \bibfield  {author} {\bibinfo {author} {\bibfnamefont {D.}~\bibnamefont
  {Knighton}},\ }\href {https://doi.org/https://doi.org/10.4324/9780203784662}
  {\emph {\bibinfo {title} {Fluvial forms and processes: a new perspective}}}\
  (\bibinfo  {publisher} {Routledge},\ \bibinfo {year} {2014})\BibitemShut
  {NoStop}%
\bibitem [{\citenamefont {Galperin}\ \emph {et~al.}(2010)\citenamefont
  {Galperin}, \citenamefont {Sukoriansky},\ and\ \citenamefont
  {Dikovskaya}}]{Galperin2010}%
  \BibitemOpen
  \bibfield  {author} {\bibinfo {author} {\bibfnamefont {B.}~\bibnamefont
  {Galperin}}, \bibinfo {author} {\bibfnamefont {S.}~\bibnamefont
  {Sukoriansky}},\ and\ \bibinfo {author} {\bibfnamefont {N.}~\bibnamefont
  {Dikovskaya}},\ }\href {https://doi.org/10.1007/s10236-010-0278-2} {\bibfield
   {journal} {\bibinfo  {journal} {Ocean Dynamics}\ }\textbf {\bibinfo {volume}
  {60}},\ \bibinfo {pages} {427} (\bibinfo {year} {2010})}\BibitemShut
  {NoStop}%
\bibitem [{\citenamefont {Davidson}(2015)}]{davidson2015turbulence}%
  \BibitemOpen
  \bibfield  {author} {\bibinfo {author} {\bibfnamefont {P.}~\bibnamefont
  {Davidson}},\ }\href@noop {} {\emph {\bibinfo {title} {Turbulence: An
  Introduction for Scientists and Engineers}}}\ (\bibinfo  {publisher} {Oxford
  University Press},\ \bibinfo {year} {2015})\BibitemShut {NoStop}%
\bibitem [{\citenamefont {Stein}\ and\ \citenamefont
  {Sabbah}(1976)}]{stein1976turbulent}%
  \BibitemOpen
  \bibfield  {author} {\bibinfo {author} {\bibfnamefont {P.~D.}\ \bibnamefont
  {Stein}}\ and\ \bibinfo {author} {\bibfnamefont {H.~N.}\ \bibnamefont
  {Sabbah}},\ }\href {https://doi.org/10.1161/01.RES.39.1.58} {\bibfield
  {journal} {\bibinfo  {journal} {Circulation Research}\ }\textbf {\bibinfo
  {volume} {39}},\ \bibinfo {pages} {58} (\bibinfo {year} {1976})}\BibitemShut
  {NoStop}%
\bibitem [{\citenamefont {Tabeling}(2002)}]{Tabeling2002}%
  \BibitemOpen
  \bibfield  {author} {\bibinfo {author} {\bibfnamefont {P.}~\bibnamefont
  {Tabeling}},\ }\href {https://doi.org/10.1016/S0370-1573(01)00064-3}
  {\bibfield  {journal} {\bibinfo  {journal} {Physics Reports}\ }\textbf
  {\bibinfo {volume} {362}},\ \bibinfo {pages} {1} (\bibinfo {year}
  {2002})}\BibitemShut {NoStop}%
\bibitem [{\citenamefont {Boffetta}\ and\ \citenamefont
  {Ecke}(2012)}]{Boffetta2012}%
  \BibitemOpen
  \bibfield  {author} {\bibinfo {author} {\bibfnamefont {G.}~\bibnamefont
  {Boffetta}}\ and\ \bibinfo {author} {\bibfnamefont {R.~E.}\ \bibnamefont
  {Ecke}},\ }\href {https://doi.org/10.1146/annurev-fluid-120710-101240}
  {\bibfield  {journal} {\bibinfo  {journal} {Annual Review of Fluid
  Mechanics}\ }\textbf {\bibinfo {volume} {44}},\ \bibinfo {pages} {427}
  (\bibinfo {year} {2012})}\BibitemShut {NoStop}%
\bibitem [{\citenamefont {Kolmogorov}(1941)}]{Kolmogorov1941}%
  \BibitemOpen
  \bibfield  {author} {\bibinfo {author} {\bibfnamefont {A.}~\bibnamefont
  {Kolmogorov}},\ }\href@noop {} {\bibfield  {journal} {\bibinfo  {journal}
  {Dokl. Akad. Nauk SSSR}\ }\textbf {\bibinfo {volume} {30}},\ \bibinfo {pages}
  {301} (\bibinfo {year} {1941})}\BibitemShut {NoStop}%
\bibitem [{\citenamefont {Sommeria}\ and\ \citenamefont
  {Moreau}(1982)}]{Sommeria1982}%
  \BibitemOpen
  \bibfield  {author} {\bibinfo {author} {\bibfnamefont {J.}~\bibnamefont
  {Sommeria}}\ and\ \bibinfo {author} {\bibfnamefont {R.}~\bibnamefont
  {Moreau}},\ }\href {https://doi.org/10.1017/S0022112082001177} {\bibfield
  {journal} {\bibinfo  {journal} {Journal of Fluid Mechanics}\ }\textbf
  {\bibinfo {volume} {118}},\ \bibinfo {pages} {507–} (\bibinfo {year}
  {1982})}\BibitemShut {NoStop}%
\bibitem [{\citenamefont {Paret}\ and\ \citenamefont
  {Tabeling}(1997)}]{ParetTabeling1997}%
  \BibitemOpen
  \bibfield  {author} {\bibinfo {author} {\bibfnamefont {J.}~\bibnamefont
  {Paret}}\ and\ \bibinfo {author} {\bibfnamefont {P.}~\bibnamefont
  {Tabeling}},\ }\href {https://doi.org/10.1103/PhysRevLett.79.4162} {\bibfield
   {journal} {\bibinfo  {journal} {Phys. Rev. Lett.}\ }\textbf {\bibinfo
  {volume} {79}},\ \bibinfo {pages} {4162} (\bibinfo {year}
  {1997})}\BibitemShut {NoStop}%
\bibitem [{\citenamefont {von Kameke}\ \emph {et~al.}(2011)\citenamefont {von
  Kameke}, \citenamefont {Huhn}, \citenamefont {Fern\'andez-Garc\'{\i}a},
  \citenamefont {Mu\~nuzuri},\ and\ \citenamefont
  {P\'erez-Mu\~nuzuri}}]{Kameke2011}%
  \BibitemOpen
  \bibfield  {author} {\bibinfo {author} {\bibfnamefont {A.}~\bibnamefont {von
  Kameke}}, \bibinfo {author} {\bibfnamefont {F.}~\bibnamefont {Huhn}},
  \bibinfo {author} {\bibfnamefont {G.}~\bibnamefont
  {Fern\'andez-Garc\'{\i}a}}, \bibinfo {author} {\bibfnamefont {A.~P.}\
  \bibnamefont {Mu\~nuzuri}},\ and\ \bibinfo {author} {\bibfnamefont
  {V.}~\bibnamefont {P\'erez-Mu\~nuzuri}},\ }\href
  {https://doi.org/10.1103/PhysRevLett.107.074502} {\bibfield  {journal}
  {\bibinfo  {journal} {Phys. Rev. Lett.}\ }\textbf {\bibinfo {volume} {107}},\
  \bibinfo {pages} {074502} (\bibinfo {year} {2011})}\BibitemShut {NoStop}%
\bibitem [{\citenamefont {Marcus}(1993)}]{Marcus1993}%
  \BibitemOpen
  \bibfield  {author} {\bibinfo {author} {\bibfnamefont {P.~S.}\ \bibnamefont
  {Marcus}},\ }\href {https://doi.org/10.1146/annurev.aa.31.090193.002515}
  {\bibfield  {journal} {\bibinfo  {journal} {Annual Review of Astronomy and
  Astrophysics}\ }\textbf {\bibinfo {volume} {31}},\ \bibinfo {pages} {523}
  (\bibinfo {year} {1993})}\BibitemShut {NoStop}%
\bibitem [{\citenamefont {Kraichnan}(1967)}]{Kraichnan1967}%
  \BibitemOpen
  \bibfield  {author} {\bibinfo {author} {\bibfnamefont {R.~H.}\ \bibnamefont
  {Kraichnan}},\ }\href {https://doi.org/10.1063/1.1762301} {\bibfield
  {journal} {\bibinfo  {journal} {The Physics of Fluids}\ }\textbf {\bibinfo
  {volume} {10}},\ \bibinfo {pages} {1417} (\bibinfo {year}
  {1967})}\BibitemShut {NoStop}%
\bibitem [{\citenamefont {Ferziger}\ and\ \citenamefont
  {Peri{\'c}}(2002)}]{ferziger2002computational}%
  \BibitemOpen
  \bibfield  {author} {\bibinfo {author} {\bibfnamefont {J.~H.}\ \bibnamefont
  {Ferziger}}\ and\ \bibinfo {author} {\bibfnamefont {M.}~\bibnamefont
  {Peri{\'c}}},\ }\href@noop {} {\emph {\bibinfo {title} {Computational Methods
  for Fluid Dynamics}}},\ Vol.\ \bibinfo {volume} {586}\ (\bibinfo  {publisher}
  {Springer},\ \bibinfo {year} {2002})\BibitemShut {NoStop}%
\bibitem [{\citenamefont {Fletcher}(2012)}]{fletcher2012computational}%
  \BibitemOpen
  \bibfield  {author} {\bibinfo {author} {\bibfnamefont {C.~A.}\ \bibnamefont
  {Fletcher}},\ }\href@noop {} {\emph {\bibinfo {title} {Computational
  Techniques for Fluid Dynamics 2: Specific techniques for different flow
  categories}}}\ (\bibinfo  {publisher} {Springer Science \& Business Media},\
  \bibinfo {year} {2012})\BibitemShut {NoStop}%
\bibitem [{\citenamefont {Doran}(2013)}]{DORAN2013201}%
  \BibitemOpen
  \bibfield  {author} {\bibinfo {author} {\bibfnamefont {P.~M.}\ \bibnamefont
  {Doran}},\ }in\ \href
  {https://doi.org/https://doi.org/10.1016/B978-0-12-220851-5.00007-1} {\emph
  {\bibinfo {booktitle} {Bioprocess Engineering Principles (Second
  Edition)}}},\ \bibinfo {editor} {edited by\ \bibinfo {editor} {\bibfnamefont
  {P.~M.}\ \bibnamefont {Doran}}}\ (\bibinfo  {publisher} {Academic Press},\
  \bibinfo {address} {London},\ \bibinfo {year} {2013})\ \bibinfo {edition}
  {second edition}\ ed.,\ pp.\ \bibinfo {pages} {201--254}\BibitemShut
  {NoStop}%
\bibitem [{\citenamefont {Wilcox}\ \emph {et~al.}(1998)\citenamefont {Wilcox}
  \emph {et~al.}}]{wilcox1998turbulence}%
  \BibitemOpen
  \bibfield  {author} {\bibinfo {author} {\bibfnamefont {D.~C.}\ \bibnamefont
  {Wilcox}} \emph {et~al.},\ }\href@noop {} {\emph {\bibinfo {title}
  {Turbulence Modeling for CFD}}},\ Vol.~\bibinfo {volume} {2}\ (\bibinfo
  {publisher} {DCW Industries La Canada, CA},\ \bibinfo {year}
  {1998})\BibitemShut {NoStop}%
\bibitem [{\citenamefont {Succi}(2001)}]{succi2001lattice}%
  \BibitemOpen
  \bibfield  {author} {\bibinfo {author} {\bibfnamefont {S.}~\bibnamefont
  {Succi}},\ }\href@noop {} {\emph {\bibinfo {title} {The Lattice {Boltzmann}
  Equation for Fluid Dynamics and Beyond}}}\ (\bibinfo  {publisher} {Clarendon
  Press},\ \bibinfo {address} {Oxford},\ \bibinfo {year} {2001})\BibitemShut
  {NoStop}%
\bibitem [{\citenamefont {Kr{\"u}ger}\ \emph {et~al.}(2017)\citenamefont
  {Kr{\"u}ger}, \citenamefont {Kusumaatmaja}, \citenamefont {Kuzmin},
  \citenamefont {Shardt}, \citenamefont {Silva},\ and\ \citenamefont
  {Viggen}}]{kruger2017lattice}%
  \BibitemOpen
  \bibfield  {author} {\bibinfo {author} {\bibfnamefont {T.}~\bibnamefont
  {Kr{\"u}ger}}, \bibinfo {author} {\bibfnamefont {H.}~\bibnamefont
  {Kusumaatmaja}}, \bibinfo {author} {\bibfnamefont {A.}~\bibnamefont
  {Kuzmin}}, \bibinfo {author} {\bibfnamefont {O.}~\bibnamefont {Shardt}},
  \bibinfo {author} {\bibfnamefont {G.}~\bibnamefont {Silva}},\ and\ \bibinfo
  {author} {\bibfnamefont {E.~M.}\ \bibnamefont {Viggen}},\ }\href@noop {}
  {\emph {\bibinfo {title} {The Lattice Boltzmann Method}}},\ Vol.~\bibinfo
  {volume} {10}\ (\bibinfo  {publisher} {Springer-Verlag},\ \bibinfo {address}
  {London},\ \bibinfo {year} {2017})\BibitemShut {NoStop}%
\bibitem [{\citenamefont {Mohamad}(2019)}]{mohamad2019}%
  \BibitemOpen
  \bibfield  {author} {\bibinfo {author} {\bibfnamefont {A.}~\bibnamefont
  {Mohamad}},\ }\href {https://doi.org/10.1007/978-1-4471-7423-3} {\emph
  {\bibinfo {title} {Lattice Boltzmann Method}}}\ (\bibinfo  {publisher}
  {Springer-Verlag},\ \bibinfo {address} {London},\ \bibinfo {year}
  {2019})\BibitemShut {NoStop}%
\bibitem [{\citenamefont {Aidun}\ \emph {et~al.}(1998)\citenamefont {Aidun},
  \citenamefont {Lu},\ and\ \citenamefont {Ding}}]{Aidun1998}%
  \BibitemOpen
  \bibfield  {author} {\bibinfo {author} {\bibfnamefont {C.}~\bibnamefont
  {Aidun}}, \bibinfo {author} {\bibfnamefont {Y.}~\bibnamefont {Lu}},\ and\
  \bibinfo {author} {\bibfnamefont {E.}~\bibnamefont {Ding}},\ }\href
  {https://doi.org/10.1017/S0022112098002493} {\bibfield  {journal} {\bibinfo
  {journal} {Journal of Fluid Mechanics}\ }\textbf {\bibinfo {volume} {373}},\
  \bibinfo {pages} {287–311} (\bibinfo {year} {1998})}\BibitemShut {NoStop}%
\bibitem [{\citenamefont {Bhatnagar}\ \emph {et~al.}(1954)\citenamefont
  {Bhatnagar}, \citenamefont {Gross},\ and\ \citenamefont
  {Krook}}]{PhysRev.94.511}%
  \BibitemOpen
  \bibfield  {author} {\bibinfo {author} {\bibfnamefont {P.~L.}\ \bibnamefont
  {Bhatnagar}}, \bibinfo {author} {\bibfnamefont {E.~P.}\ \bibnamefont
  {Gross}},\ and\ \bibinfo {author} {\bibfnamefont {M.}~\bibnamefont {Krook}},\
  }\href {https://doi.org/10.1103/PhysRev.94.511} {\bibfield  {journal}
  {\bibinfo  {journal} {Phys. Rev.}\ }\textbf {\bibinfo {volume} {94}},\
  \bibinfo {pages} {511} (\bibinfo {year} {1954})}\BibitemShut {NoStop}%
\bibitem [{\citenamefont {Qian}\ \emph {et~al.}(1992)\citenamefont {Qian},
  \citenamefont {D'Humières},\ and\ \citenamefont {Lallemand}}]{Qian_1992}%
  \BibitemOpen
  \bibfield  {author} {\bibinfo {author} {\bibfnamefont {Y.~H.}\ \bibnamefont
  {Qian}}, \bibinfo {author} {\bibfnamefont {D.}~\bibnamefont {D'Humières}},\
  and\ \bibinfo {author} {\bibfnamefont {P.}~\bibnamefont {Lallemand}},\ }\href
  {https://doi.org/10.1209/0295-5075/17/6/001} {\bibfield  {journal} {\bibinfo
  {journal} {Europhysics Letters}\ }\textbf {\bibinfo {volume} {17}},\ \bibinfo
  {pages} {479} (\bibinfo {year} {1992})}\BibitemShut {NoStop}%
\bibitem [{\citenamefont {Smagorinsky}(1963)}]{Smagorinsky1963}%
  \BibitemOpen
  \bibfield  {author} {\bibinfo {author} {\bibfnamefont {J.}~\bibnamefont
  {Smagorinsky}},\ }\href
  {https://doi.org/10.1175/1520-0493(1963)091<0099:GCEWTP>2.3.CO;2} {\bibfield
  {journal} {\bibinfo  {journal} {Monthly Weather Review}\ }\textbf {\bibinfo
  {volume} {91}},\ \bibinfo {pages} {99} (\bibinfo {year} {1963})}\BibitemShut
  {NoStop}%
\bibitem [{\citenamefont {Zou}\ and\ \citenamefont
  {He}(1997)}]{zou1997pressure}%
  \BibitemOpen
  \bibfield  {author} {\bibinfo {author} {\bibfnamefont {Q.}~\bibnamefont
  {Zou}}\ and\ \bibinfo {author} {\bibfnamefont {X.}~\bibnamefont {He}},\
  }\href {https://doi.org/10.1063/1.869307} {\bibfield  {journal} {\bibinfo
  {journal} {Physics of fluids}\ }\textbf {\bibinfo {volume} {9}},\ \bibinfo
  {pages} {1591} (\bibinfo {year} {1997})}\BibitemShut {NoStop}%
\bibitem [{\citenamefont {Kundu}\ \emph {et~al.}(2024)\citenamefont {Kundu},
  \citenamefont {Cohen}, \citenamefont {Dowling},\ and\ \citenamefont
  {Capecelatro}}]{kundu2024fluid}%
  \BibitemOpen
  \bibfield  {author} {\bibinfo {author} {\bibfnamefont {P.~K.}\ \bibnamefont
  {Kundu}}, \bibinfo {author} {\bibfnamefont {I.~M.}\ \bibnamefont {Cohen}},
  \bibinfo {author} {\bibfnamefont {D.~R.}\ \bibnamefont {Dowling}},\ and\
  \bibinfo {author} {\bibfnamefont {J.}~\bibnamefont {Capecelatro}},\
  }\href@noop {} {\emph {\bibinfo {title} {Fluid Mechanics}}}\ (\bibinfo
  {publisher} {Elsevier},\ \bibinfo {year} {2024})\BibitemShut {NoStop}%
\bibitem [{\citenamefont {Lilly}(1972)}]{lilly1972numerical}%
  \BibitemOpen
  \bibfield  {author} {\bibinfo {author} {\bibfnamefont {D.~K.}\ \bibnamefont
  {Lilly}},\ }\href {https://doi.org/10.1080/03091927208236087} {\bibfield
  {journal} {\bibinfo  {journal} {Geophysical Fluid Dynamics}\ }\textbf
  {\bibinfo {volume} {4}},\ \bibinfo {pages} {1} (\bibinfo {year}
  {1972})}\BibitemShut {NoStop}%
\bibitem [{\citenamefont {Herring}\ \emph {et~al.}(1974)\citenamefont
  {Herring}, \citenamefont {Orszag}, \citenamefont {Kraichnan},\ and\
  \citenamefont {Fox}}]{herring1974decay}%
  \BibitemOpen
  \bibfield  {author} {\bibinfo {author} {\bibfnamefont {J.~R.}\ \bibnamefont
  {Herring}}, \bibinfo {author} {\bibfnamefont {S.~A.}\ \bibnamefont {Orszag}},
  \bibinfo {author} {\bibfnamefont {R.~H.}\ \bibnamefont {Kraichnan}},\ and\
  \bibinfo {author} {\bibfnamefont {D.~G.}\ \bibnamefont {Fox}},\ }\href
  {https://doi.org/10.1017/S0022112074000280} {\bibfield  {journal} {\bibinfo
  {journal} {Journal of Fluid Mechanics}\ }\textbf {\bibinfo {volume} {66}},\
  \bibinfo {pages} {417} (\bibinfo {year} {1974})}\BibitemShut {NoStop}%
\bibitem [{\citenamefont {Mcwilliams}(1984)}]{Mcwilliams_1984}%
  \BibitemOpen
  \bibfield  {author} {\bibinfo {author} {\bibfnamefont {J.~C.}\ \bibnamefont
  {Mcwilliams}},\ }\href {https://doi.org/10.1017/S0022112084001750} {\bibfield
   {journal} {\bibinfo  {journal} {Journal of Fluid Mechanics}\ }\textbf
  {\bibinfo {volume} {146}},\ \bibinfo {pages} {21–43} (\bibinfo {year}
  {1984})}\BibitemShut {NoStop}%
\end{thebibliography}
